\documentclass{aa}
\usepackage{graphicx}
\usepackage{txfonts}
\usepackage[figuresright]{rotating}
\begin{document}
\title{Proper motions and membership probabilities of stars in the region of 
globular cluster NGC~6809\thanks{Based   on
   observations with the MPG/ESO  2.2m and ESO/VLT telescopes, located
   at  La Silla  and  Paranal Observatory,  Chile,  under DDT  programs
   077.D-0372(A), 163.O-0741(C) and the archive material.}}

\author{ Devesh\ P.\ Sariya\inst{1},
          R.\ K.\ S.\ Yadav\inst{1},
          A.\ Bellini\inst{2}.}
     
  \institute{
  Aryabhatta Research Institute of Observational Sciences, 
  Manora Peak, Nainital 263 129, India \\
  \email{devesh; rkant@aries.res.in}
  \and
  Dipartimento di Astronomia, Universita di Padova, Vicolo dell' Osservatorio 2, I-35122 Padova, I, EU\\
  \email{andrea.bellini@unipd.it}
 }

%\date{Received 13 December 2007 / Accepted 25 February 2008}

% \abstract{}{}{}{}{} 
% 5 {} token are mandatory 
\abstract
    {NGC 6809 is a luminous metal-poor halo globular cluster that  
     is relatively easy to study due to its proximity and low 
     concentration. Because of its high Galactic latitude ($b = -23^\circ$), 
     interstellar reddening and contamination is not very high. }
%    % aims heading (mandatory)
    { We aim to determine the relative proper 
      motion and membership probability of the stars in the wide area of 
      globular cluster NGC 6809. To target cluster members reliably during 
      spectroscopic surveys and both spatial and radial distributions in 
      the cluster outskirts without including field stars, a good proper 
      motion and membership probability catalogue of NGC 6809 is required.}  
%    % methods heading (mandatory)
     {The archival data of two epochs with a time-base line of 7.1 years 
      have been collected with Wide Field Imager (WFI) mounted on the 2.2m 
      MPG/ESO telescope. The CCD images of both epochs 
      have been reduced using the astrometric techniques as described 
      in Anderson et al. (2006). The calibrated $UBVI$ magnitudes have
      been derived using Stetson's secondary standard stars.} 
%    % results heading (mandatory)
     {We derived the relative proper motion and membership probabilities
     for $\sim$ 12600 stars in the field of globular cluster NGC 6809.
     The measurement error in proper motions for the stars of $V \sim 17$ mag
     is 2.0  mas~yr$^{-1}$, gradually increasing up to  $\sim$3 mas~yr$^{-1}$ 
     at $V=20$ mag. We also provide the membership probability for the published        
     different types of sources in NGC 6809. An electronic 
     catalogue \thanks{Full Table 5 is only available in electronic form 
     at the CDS via anonymous ftp to cdsarc.u-strasbg.fr(130.79.128.5) 
     or via http://cdsweb.u-strasbg.fr/cgi-bin/qcat?J/A+A/} with 
     proper motion and membership probability for the stars will
     be available to the astronomical community.}
    {}
%    %
    
    \titlerunning{Astrometry of NGC~6809}
    \authorrunning{Devesh P. Sariya et al.} 
    \maketitle

%________________________________________________________________
%
%
\section{Introduction}
\label{INTR}
%________________________________________________________________
%
Globular clusters have long been used to study the structure and formation
of our Galaxy. NGC~6809 (M~55) is a sparse, metal-poor globular cluster whose
proximity ($\sim$ 5 kpc) makes it an excellent target for an in-depth study
of its stellar population.

NGC~6809 harbours several interesting objects.  Bassa et al.\ (2008) found 16
$X$-ray sources within the half-mass radius (2\arcmin.89) of NGC~6809, of
which eight or nine are expected to be background sources. On the basis of optical
counterparts, these authors identified three sources related to the cluster. The brightest
$X$-ray source of this cluster is classified as a dwarf nova. Blue stragglers
in NGC 6809 have been studied by Lanzoni et al. (2007) and exhibit a bimodal
radial distribution with a central peak, a broad minimum at intermediate radii,
and an upturn outwards. A detailed study about the evolved stars, including
asymptotic giant branch, horizontal branch, and upper red giant branch (RGB) stars
was presented by Vargas et al. (2007).

A proper-motion (PM) study of NGC~6809 was conducted by Dinescu et al. (1999)
using photographic plates. The authors determined an absolute proper motion
($\mu_{\alpha}cos\delta=-1.57\pm0.62$ mas yr$^{-1}$, $\mu_{\delta}=-10.14\pm0.64$ mas yr$^{-1}$)
using a sample of $\sim$ 600 cluster stars brighter than $V\sim16$ photographic
magnitude and background galaxies as a reference. Recently, a proper motion
study of NGC~6809 was performed by Zloczewski et al. (2011) (hereafter, Zl11).
They determined membership probabilities
for 16,645 stars in the central part of the cluster
($8\farcm83$ $\times$ $8\farcm83$) and found an absolute proper motion
$\mu_{\alpha}cos\delta=-3.31\pm0.10$ mas yr$^{-1}$ and $\mu_{\delta}=-9.14\pm0.15$ mas yr$^{-1}$.

Despite the extensive photometric studies of this cluster, there is a lack of studies that
provide proper motions and membership probabilities in the wide field region for NGC 6809.
Wide-field images allow us to map completely any open or globular cluster
in our Galaxy and its tidal trails and allow us to obtain accurate photometry for an 
enormous number of stars.
In combination in photometric data, membership information is very useful to keep the
field star contamination to a minimum. The archival wide-field multi-epoch observations taken 
with the WFI@{\bf 2.2}m telescope offer new opportunities to derive precise PMs with
only a few years of time span, deeper by several magnitudes than previous photographic
surveys (Anderson et al. 2006; Yadav et al. 2008; Bellini et al. 2009).

The main purpose of the present study is to determine accurate relative PMs and
membership probabilities for stars brighter than $V\sim20$ mag in the wide area of NGC 6809.
Membership probabilities of different sources in NGC 6809 are also discussed.
Fundamental parameters of the cluster taken from Harris (1996) are listed in
Table~\ref{par}. The PMs, membership probabilities, and photometric $U,B,V$ and
$I$ magnitudes are provided to the astronomical community for follow-up
studies. Our membership-probability catalogue is in wide field ($26\arcmin \times 22\arcmin$) and
contains $UBVI$ magnitudes, while Zl11 have provided membership probabilities only in $V$ magnitude
and in the central area of $8\farcm83$ $\times$ $8\farcm83$.

The structure of the article is as follows. Data taken for the present study, their
reduction procedures and comparison of photometric and astrometric data are described in
Sect.~\ref{OBS}, where we also determine PMs and differential chromatic refraction.
In Sect.~\ref{MP} we present the cluster membership analysis. In Sect.~\ref{app}
we use our catalogue to confirm the membership of previously found variables,
blue stragglers, and X-ray sources. Finally, Sect.~\ref{catl} describes the catalogue while 
Sect.~\ref{con} represents the conclusions of the present study.

\begin{table}
\caption{Fundamental parameters of NGC~6809 taken from Harris (1996). }
\centering
\label{par}
\begin{tabular}{cc}
\hline\hline Parameters & Values \\ \hline
%\hline\hline  $\alpha_{2000}$ & $\delta_{2000}$ & $l$ & $b$ & $\rm [Fe/H]$ & $E(B-V)$ & $(m-M)$ \\ \hline
%19$^{\rm h}$ 39$^{\rm m}$ 59.$^{\rm s}$4 & $-30^\circ$ 57$\arcmin$ 44$\arcsec$& $8\fdg80$ & $-23\fdg27$ & $-1.81 $ & 0.08 &13.87 \\
$\alpha$(J2000)  &     19$^{\rm h}$ 39$^{\rm m}$ 59.$^{\rm s}$4  \\
$\delta$(J2000)  & $-30^\circ$ 57$\arcmin$ 44$\arcsec$    \\
$l$              & $8\fdg80$                \\
$b$              & $-23\fdg27$              \\
$\rm [Fe/H]$     & $-1.81 $           \\
$E(B-V)  $       & 0.08   \\
$(m-M)$          & 13.87 \\
\hline
\end{tabular}
\end{table}

%__________________________________________________ % %

%________________________________________________________________
%
%

%%%%%%%%%%%%%%%%

\section{Archival data and reductions}
\label{OBS}
%________________________________________________________________
%
Proper motions of the stars in the cluster region were computed using archive
WFI@{\bf 2.2}m $V$ filter images\footnote{http://archive.eso.org/eso/eso\_archive\_main.html.}
taken in 1999 (first epoch) and in 2006 (second epoch). The 1999 observing run
consists of two images in $B$ filter, taken in May, while four
images in $V$ and eight images in $I$ were taken in
July. Six $U$ -band images and six $V$-band images were
acquired in August, 2006.

The WFI@{\bf 2.2}m consists of eight 2048$\times$4096 EEV CCDs with
$0\farcs238$ pixel$^{-1}$ resulting in a total field-of-view
34$\arcmin \times 33\arcmin$. The observational log with epoch information
is listed in Table~\ref{log}. Images used in the analysis were taken under similar
seeing ($\sim 1^{\prime\prime}$) and airmass conditions.
Long and short exposures were acquired to map the brighter and
fainter stars of the cluster. The data used in this article were not collected
with the aim of proper motion study.

%%%%%%%%%%%%%%%%%%%%%%%%%%%%%%%%%%%%%%%%%%%%%%%%%%%%%%%%%%%%%%%%%%%%%%%%%%%%%%%%%%%%%%%5
\begin{table}
%\begin{center}
\caption{Description of the WFI@{\bf 2.2}m data sets.
  The first epoch data were observed in May and July, 1999, 
  while second epoch data were observed in August, 2006.}
\label{log}
\begin{tabular}{cccc}
\hline
\hline
Filters      &  Exposure Time & Seeing & Airmass \\
&(in seconds)&& \\
\hline
\multicolumn{4}{c}{1999 May--July} \\
$B$&2$\times$240&1$''$.0&1.0     \\
$V/89$ (First epoch)&4$\times$200&1$''$.1&1.1     \\
$I/lwp$&7$\times$150; 1$\times$200&1$''$.0&1.1     \\
\multicolumn{4}{c}{2006 August} \\
$U/50_{-}ESO877$&6$\times$300&1$''$.1&1.0     \\
$V/89$ (Second epoch)&6$\times$40&1$''$.0&1.0         \\
\hline
\end{tabular}
%\end{center}
\end{table}
%%%%%%%%%%%%%%%%

%%%%%%%%%%%%%%%%
%%%%%%%%%%%%%%%%
\begin{figure}
\centering
\includegraphics[width=8.5cm]{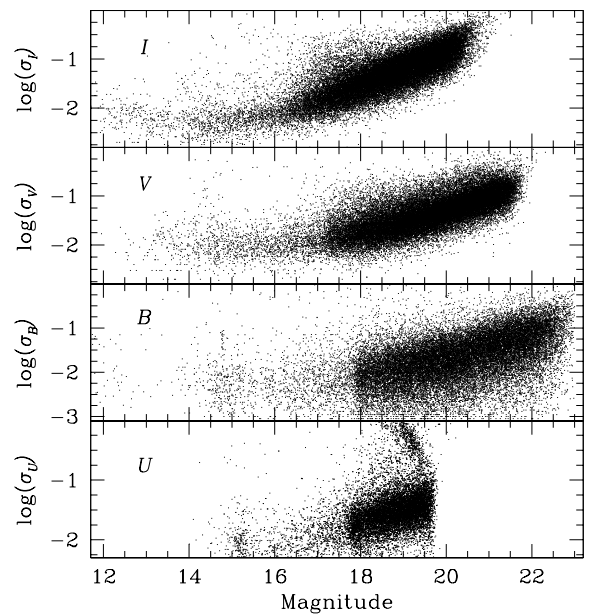}
\caption{ Plot of the rms logarithm of the residuals around the mean magnitudes
in $U,B,V$ and $I$ as a function of their magnitudes.}
\label{error_mag}
\end{figure}
%%%%%%%%%%%%%%%%%%%%%%%%%%%%%

\subsection{Astromeric and photometric reductions}

The reduction procedures described in Anderson et al. (2006, Paper~I)
were adopted for the WFI@{\bf 2.2}m CCD images.
This includes de-biasing, flat-fielding, and
correction for cosmic rays. In Paper~I we showed that the WFI@{\bf 2.2}m PSF
changes significantly with position on the detector. Because of this, an array
of empirical point spread functions (PSFs) were constructed for each image to
obtain the positions and fluxes of the objects. These PSFs are saved in a look-up
table on a very fine grid. To select suitable stars for the PSFs, we
developed an automatic code (see Paper~I). An iterative process is designed
to work from the brightest down to the faintest stars and find their precise
position and instrumental flux for $U$, $B$, $V$ and $I$ exposures.

%%%%%%%%%%%%%%%%
\begin{figure}
\centering
\includegraphics[width=8.5cm]{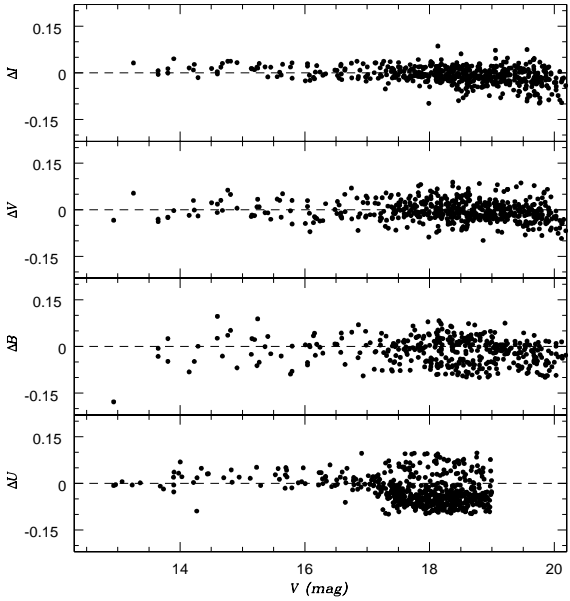}
\caption{Star-by-star comparison of our photometry with standard stars.
The $BVI$ magnitudes were compared with
the Stetson catalogue, while the
%\footnote{http://cadcwww.hia.nrc.ca/standards/}while, for 
$U$ filter was compared with data taken from
Kaluzny et al. (2005).
No clear systematics are present, which proves the reliability of our
calibration procedures.}
\label{comp_stet}
\end{figure}
%%%%%%%%%%%%%%%%

In Paper~I we showed that WFI@{\bf 2.2}m has a large geometric distortion,
i.e. the pixel scale is changing across the field of view. To derive the
correction for the geometric distortion, we parametrized the
distortion solution by a look-up table of corrections
for each chip that covered 2048$\times$4096 pixel each, sampling
every 256 pixels. This resulted in a 9$\times$17 element array of
corrections for each chip. At any given location on the  detector, a
bi-linear interpolation between the four closest grid points of the
look-up table provided the corrections for the
target point. The derived look-up table may have a lower accuracy on
the edges of a field because of the way the self-calibration frames
were dithered (see Paper~I). An additional source of uncertainty is
related to a possible instability of distortions for the $WFI@{\bf 2.2}$m
reported earlier. This prompted us to use the local-transformation
method to derive PMs. Detailed descriptions about the
distortion solution are given in Paper~I.

In the local transformation approach a small set of
local reference stars is selected around each target object. It is
advantageous to use pre-selected cluster members to form a local
reference frame because of the much lower intrinsic velocity dispersion
among the cluster members.
Then, six-parameter linear transformations are used to transform the
coordinates from one frame into another, taken at different epochs.
The residuals of this transformation characterize relative
PMs convolved with measurement errors. In essence, this is
a classical ``plate pair'' method but extended to all possible
combinations of the first- and second- epoch frames. The relative PM
of a target object is an average of all displacement measurements in
its local reference frame. The last step is to estimate the
measurement errors from intra-epoch observations where PMs
have a zero contribution. A complete description of all steps leading
to PMs is given in Paper~I.

\subsubsection{Photometric calibration}

% A:
In order to transform instrumental $U$, $B$, $V$, $I$ magnitudes into the
standard Johnson-Cousin system, we used a list of secondary standard stars provided by
Stetson\footnote{http://cadcwww.hia.nrc.ca/standards/} for $B$, $V$ and $I$
and by Kaluzny et al. (2005) for $U$ filter. In total, 851 common stars were found in
Stetson's catalogue and were used to calibrate $B$, $V$ and $I$ magnitudes.
These common stars have a brightness range of 12.9$\le V \le$20.6 mag and a colour
range of $-0.5\le(V-I)\le$2.0 mag, which
coveres all stars brighter than $V=20.0$ mag. 1205 common stars brighter
than {\bf$V=19$} mag were found in the Kaluzny et al. (2005) catalogue. These stars
have a colour range of $-0.5\le(U-B)\le$1.3 mag.

We derived photometric zero-points and colour terms using the
following transformation equations:\\

$ U_{\rm std} = U_{\rm ins} + C_u*(U_{\rm ins} - B_{\rm ins}) + Z_u $

$ B_{\rm std} = B_{\rm ins} + C_b*(B_{\rm ins} - V_{\rm ins}) + Z_b $

$ V_{\rm std} = V_{\rm ins} + C_v*(V_{\rm ins} - I_{\rm ins}) + Z_v $

$ I_{\rm std} = I_{\rm ins} + C_i*(V_{\rm ins} - I_{\rm ins}) + Z_i $,\\

where the subscript ``ins'' means instrumental magnitudes and ``std''
stands for secondary standard magnitudes. $C_u, C_b, C_v$ and $C_i$ are
the colour terms while $Z_u, Z_b, Z_v$ and $Z_i$ are the global zero-points.
The quadratic colour terms are negligible. The values of colour terms
are 0.00, 0.70, 0.00, and 0.11, while the zero-points are 22.68, 24.95,
24.18, and 23.34 for $U$, $B$, $V$ and $I$ filters respectively.
Our colour terms and zero-point values are nearly consistent with those posted
on the WFI@{\bf 2.2}m
webpage\footnote{http://www.ls.eso.org/lasilla/sciops/2p2/E2p2M/WFI/zeropoints/}.

%%%%%%%%%%%%%%%%
\begin{figure}
\centering
\includegraphics[width=8.5cm]{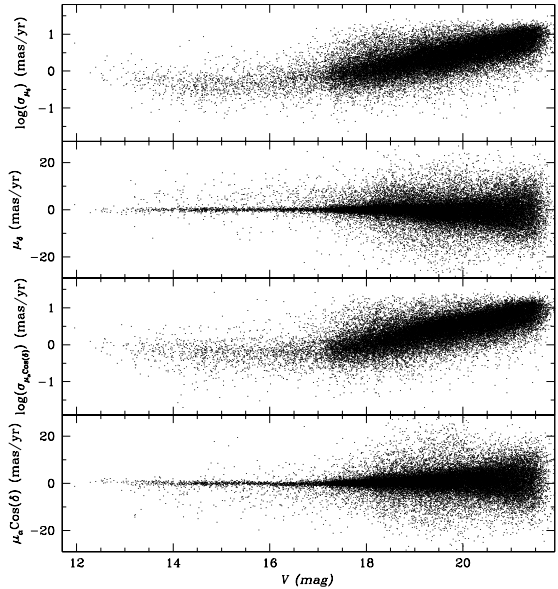}
\caption{Plots of proper motions and log of their standard deviations of the mean
versus  visual magnitude in mas~yr$^{-1}$.}
\label{error_pm}
\end{figure}
%%%%%%%%%%%%%%%%

In  Fig.~\ref{error_mag}  we show the log of rms of the residuals around the mean
magnitude for each filter as a function of the respective magnitudes.
The photometric standard deviations were computed from multiple
observations, all reduced to the common photometric reference frame in the chosen bandpass.
On average, photometric rms are better than $\sim$0.03 mag for stars brighter
than 18.0 mag in  $U, B$ and $V$ filters. Stars brighter than $I=16$ mag
have photometric rms lower than 0.01 mag, gradually increasing to 0.1 mag for $I=20.0$ mag.

Figure~\ref{comp_stet} exhibits the magnitude difference between our
calibrated $B$, $V$ and $I$ magnitudes and the Stetson secondary standards as a function
of $V$ magnitude. The difference between our calibrated $U$ magnitudes
and the Kaluzny et al. (2005) data is also shown as a function of $V$ magnitude.
There are no clear systematic trends seen in the differences with $V$ mag.

\subsubsection{Astrometric calibration}
\label{ac}

The next step is to transform $X$ and $Y$ coordinate to right
ascension (RA) and declination (Dec). The $X, Y$ raw positions of each star in
each frame were corrected for geometric distortion using the
look-up table provided in Paper I, brought into common reference frames
by means of six-parameter linear transformations and averaged.
To transform the averaged $X$ and $Y$ coordinates into RA and Dec of
J2000, we used the online digitized sky ESO catalogue in skycat software as an
absolute astrometric reference frame. Thanks to our accurate geometric-distortion solution and a
reasonable stability of the intra-chip positions, it was possible to
apply a single plate model involving linear and quadratic terms and a
small but significant cubic term in each coordinate. This solution
also absorbs effects caused by differential refraction but not its
chromatic component. The standard error of equatorial solution is
$\sim$ 100 mas in each coordinate.

\subsubsection{Proper motions}
\label{ppm}

Proper motions were computed using $V$-filter images to minimize
colour-dependent terms in our analysis. Moreover, our geometric-distortion
solution provides the lowest residuals with $V$ images (Paper~I).
A total of four images for the first epoch and six images for the second
epoch were used.

First, we selected a sample of probable cluster members using the $V$
vs. $(V-I)$ CMD. Selected stars are located on the RGB and MS in the magnitude
range 13.0$\le V \le18.0$ mag. These stars define a local reference frame
to transform the positions of a given first-epoch image into positions
of a second-epoch image. By adopting only stars on cluster sequences
whose proper motion errors $<$1.5 mas~yr$^{-1}$, we made sure that PMs are
measured relative to the bulk motion of the cluster. To minimize the effect of
uncorrected distortion residuals
we used the
local transformation approach based on the closest 25 reference stars
on the same CCD chip. No systematics larger than random errors are
visible close to the corners or edges of chips.

We iteratively removed some stars from the preliminary photometric
member list that had proper motions clearly inconsistent with cluster
membership, even though their colours placed them near the fiducial
cluster sequence.
According to Pryor \& Meylan (1993) the intrinsic velocity dispersion of
stars in the cluster NGC 6809 is 4.9 km/s. With a distance of 5.3 Kpc (Harris 1996)
the internal proper motion dispersion becomes 0.2 mas~yr$^{-1}$.
The distribution log of proper motions errors with $V$ mag is presented in Fig.
\ref{error_pm} for both coordinates. The precision of the proper
motion measurement (rms of individual measurement) is better than 2
mas yr$^{-1}$ upto 17.5 mag in $V$. Errors are gradually increasing upto 7
mas yr$^{-1}$ for $V=20$ mag.

%%%%%%%%%%%%%%%%
\begin{figure}
\centering
\includegraphics[width=8.5cm]{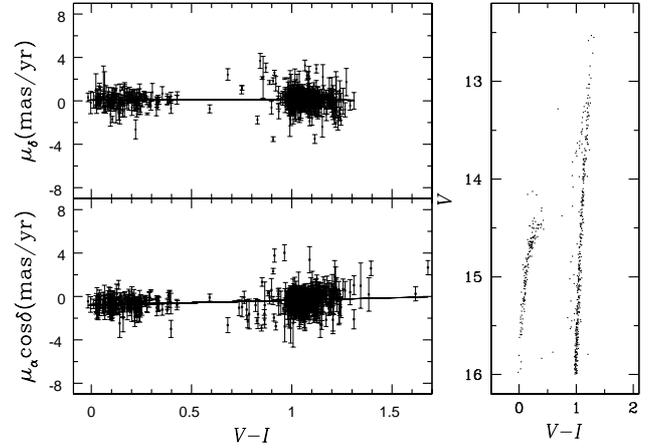}
\caption{Distribution of PMs as a function of colour for 
stars brighter than V$<$16 mag is modelled with a linear fit.}
\label{aber}
\end{figure}

%%%%%%%%%%%%%%%%

%_____________________________________________________________________________
\begin{figure*}
\centering
\includegraphics[width=\textwidth]{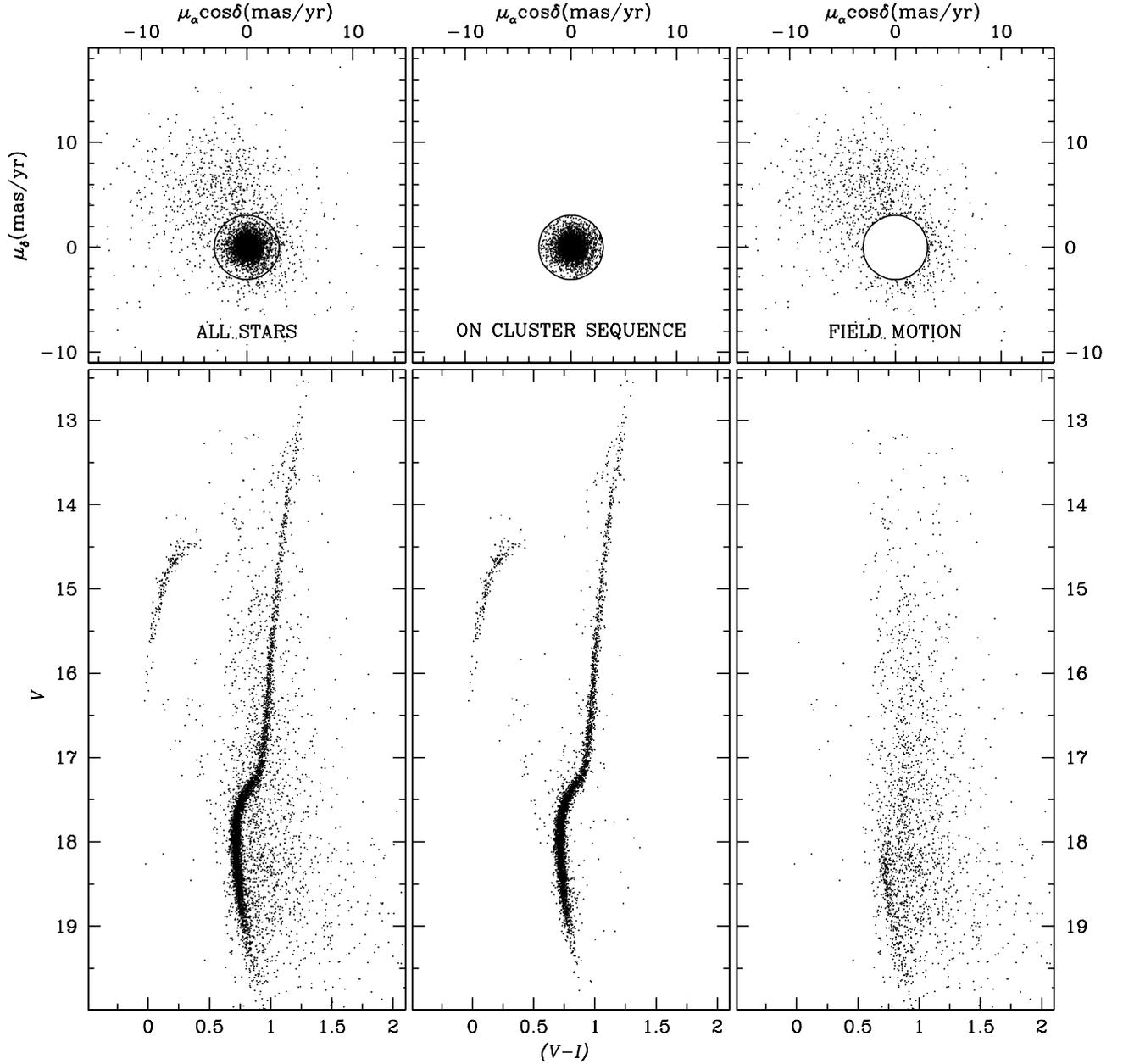}
\caption{
%       %
        {\em (Top panels)} Proper motion vector-point diagram.
        Zero point in VPD is the mean motion of cluster stars.
%       %
        {\em (Bottom panels)} Calibrated
$V$ vs. $V-I$ CMD.
%       %
        {\em (Left)} The entire sample;
%       %
        {\em (centre)} stars in VPD with proper motions within 3 mas~yr$^{-1}$
         around the cluster mean.
%       %
        {\em (Right)} Probable background/foreground field stars in the area
        of NGC~6809 studied in this paper.
%       %
        All plots show only stars with proper motion
        $\sigma$ smaller than $\sim$2.5 mas~yr$^{-1}$ in each
        coordinate.  }
%       %
\label{cmd_inst}
\end{figure*}
%________________________________________________________________

%________________________________________________________________
%
%________________________________________________________________
%
\begin{figure*}
\centering
\includegraphics[width=\textwidth]{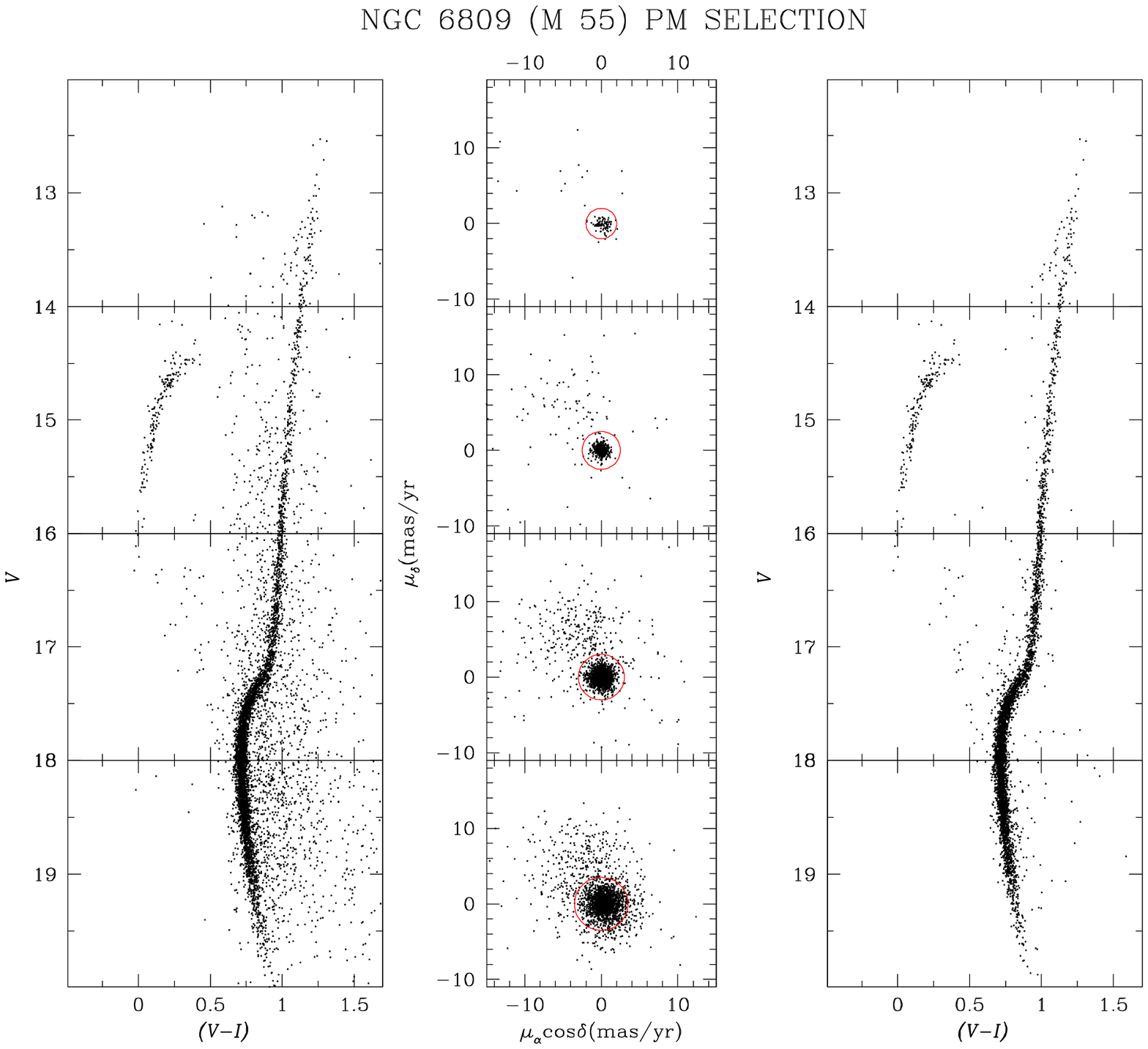}
\caption{
%       %
        {\em (Left:)} Colour-magnitude diagram for all stars whose proper motion errors increase from 1.5 mas~yr$^{-1}$ for the brightest bin to 3.5 mas~yr$^{-1}$ for the faintest one.
%       %
%       %he 
        {\em (Middle:)} Vector-point diagram for the same stars in
        corresponding magnitude intervals. A circle in each plot
        shows the adopted membership criterion.
%       % 
        {\em (Right:)}  Colour-magnitude diagram for stars assumed
        to be cluster members.
%       %
        }
%       %
%       %
\label{cmd_II}
\end{figure*}
%________________________________________________________________

\subsubsection{Differential chromatic refraction}
\label{chrom}
Because of the wavelength dependence of the refractive index of air,
a star observed with a blue filter will appear slightly higher in the
sky than the same star observed through a red filter. Unfortunately,
the data do not contain exposures suitable to properly
correct this effect. We can, however, remove possible differences in
the average differential chromatic refraction (DCR) between the two epochs.

Figure \ref{aber} shows the proper motion $\mu_{\alpha}cos(\delta)$ and $\mu_{\delta}$
as a function of $(V-I)$ colour of horizontal and giant branch stars. These selected
stars are brighter than 16$^{\bf th}$ mag in $V$ and have a proper motion error
$\le 2.5$ mas yr$^{-1}$. We see that there is a small colour-related displacement of
$\sim$$-$0.81 mas yr$^{-1}$ $(V-I)^{-1}$ in $\mu_{\alpha}cos(\delta)$ direction.
Our final proper motion data were corrected for the DCR effect by applying the colour
term correction in the proper motions.

\subsection{Cluster CMD decontamination}
\label{PM}

The vector-point diagram (VPD) of our PM measurements is shown in the top panels 
of Fig. \ref{cmd_inst} while $V$ vs. $(V-I)$ CMD in the bottom panels.
Left panels show all stars while middle and right panels show the probable cluster
members and field stars. A circle of 3 mas yr$^{-1}$ around the cluster centroid 
in VPD of proper motions defines our membership criterion. The chosen radius
is a compromise between losing cluster members with poor proper motions and
including field stars that share the cluster mean proper motion. The shape of
the cluster members' PM dispersion is round, providing that our PM
measurements are not affected by any systematics.
The right lower panel represents the CMD for field stars. A few cluster
members are also visible in this CMD because of their poorly determined
proper motions. In Fig. \ref{cmd_II}, we show the $(V-I), V$ CMD which
is binned along the magnitude axis. 
In each bin we adopted different selection criteria
to identify cluster members, which were more stringent
for stars with more reliable measurements from data of high
signal-to-noise ratio, and less restrictive for stars with less
precise measurements. The proper motion error is $<$ 1.5 mas yr$^{-1}$ 
for the brightest magnitude bin and up to 3.5 mas yr$^{-1}$ 
for the faintest bin. 
This figure shows that fainter stars have
a stronger error in proper motion.
The bright stars from the short exposures (40 sec) in epoch 2
have saturated in the long exposures (200 sec) of epoch 1.
The separation of brighter cluster members from the field stars
is clearly visible while fainter members are not clearly separated
out as seen in Fig. \ref{cmd_II}. The reason may be that proper motions
for fainter stars are not determined accurately.

%%%%%%%%%%%%%%%%
\begin{figure}
\centering
\includegraphics[width=9.0cm,height=9.0cm]{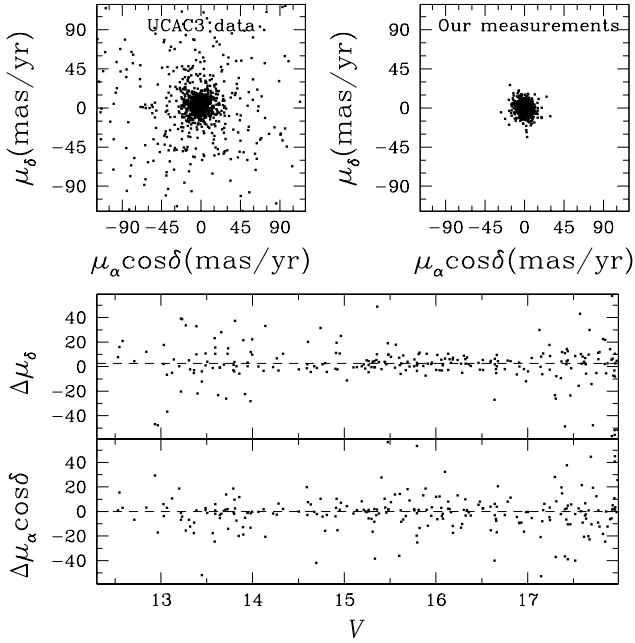}
\caption{{\it Top panels:} Vector-point diagrams of common stars
relative to the cluster mean motion for UCAC3 {\it (left)} and our catalogue {\it (right).
(Bottom panels:)} right ascension {\it(bottom)} and declination {\it(top)}
proper motion differences as a function of $V$ magnitude between UCAC3 and our
measurements. Horizontal dashed lines show the 3$\sigma$-clipped median of the
proper motion difference between UCAC3 and our data. }
\label{ucac3}
\end{figure}

%%%%%%%%%%%%%%%%
%%%%%%%%%%%%%%%%
\begin{figure}
\centering
\includegraphics[width=8.5cm,height=8.0cm]{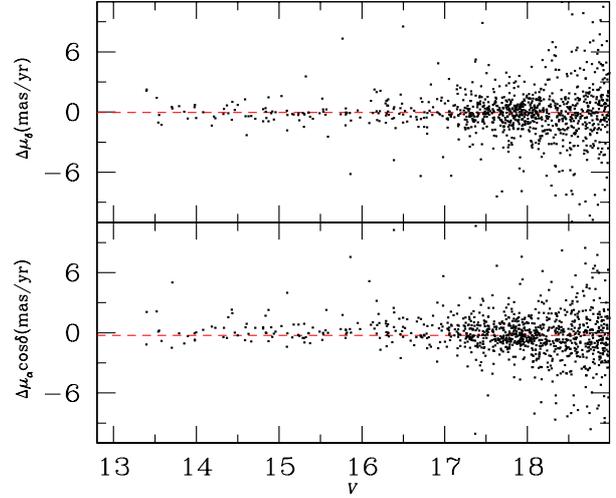}
%\vspace{-3.5cm}
\caption{Plot of proper motion differences between Zl11 and our data
with $V$ magnitude. Horizontal dashed lines show the 3$\sigma$-clipped median of the
proper motion difference between Zl11 and our data. }
\label{kamil}
\end{figure}
%%%%%%%%%%%%%%%%%%%

\subsection{Comparison with previous proper motion studies}

Absolute proper motion can be determined using galaxies observed in the
field of NGC~6809. Few background faint galaxies are visible in our images,
but they are not suitable to determine absolute proper motions because 
of the large error in their proper motions.

The absolute proper motions of stars brighter than $V=18$ mag are
available in the UCAC3 catalogue (Zacharias et al. 2010). There are 310
common stars found to be brighter than $V=18$ mag.
To compare our proper motions with UCAC3, we changed the
UCAC3 proper motions to relative proper motions. For this, we
subtracted the absolute proper motion calculated by Dinescu et al. (1999),
($\mu_{\alpha}cos{\delta} = - 1.57$ mas yr$^{-1}$, $\mu_{\delta}= -10.14$ mas yr$^{-1}$)
from the individual proper motion of UCAC3. We considered the value of the absolute proper
motion for the cluster by Dinescu et al. (1999) because it is more precise than the derived
mean absolute proper motion using UCAC3 stars. Figure \ref{ucac3} shows
the comparison of our PMs with those of the UCAC3 catalogue. The
top-left panel shows the VPD of UCAC3 stars while the top-right panel
shows the VPD of our measurements. A concentration of stars around (0,
0) mas yr$^{-1}$ is seen in both VPDs. Our proper motions distribution is
tighter than the UCAC3 distribution. This is because our data are
more precise than the UCAC3 data.

In the lower panels of Fig. \ref{ucac3} we show the difference ($\Delta$) in
$\mu_{\alpha}cos\delta$ and $\mu_{\delta}$ in the sense of our minus the UCAC3 difference as
a function of $V$ magnitude. There is no systematic trend in the differences with
magnitude. The 3$\sigma$ clipped median of differences are
$-0.10 (\sigma = 3.70)$ mas yr$^{-1}$ and $2.60 (\sigma = 4.45)$ mas yr$^{-1}$
in $\mu_{\alpha}cos\delta$ and $\mu_{\delta}$.

Fig. \ref{kamil} shows the difference between our proper motions and Zl11
plotted with $V$ magnitude. The 3$\sigma$ clipped median of
differences are
$-0.28 (\sigma = 0.86)$ mas yr$^{-1}$ and $-0.02 (\sigma = 0.82)$ mas yr$^{-1}$.
Clearly, our measurements are consistent with the Zl11 data for $V\le 19$ mag.

%

%________________________________________________________________
%
\section{Determination of membership probability}
\label{MP}
%________________________________________________________________
%
Accurate membership determination is essential for additional astrophysical
studies of cluster. The fundamental mathematical model set up by
Vasilevskis et al. (1958) and the technique based upon the maximum likelihood
principle developed by Sanders (1971) for membership determination have
since been continuously refined.

An improved method for membership determination of stars in clusters based
on proper motions with different observed precisions was developed by Stetson (1980)
and Zhao \& He (1990). Zhao \& Shao (1994) then added the
correlation coefficient of the field star distribution to the
set of parameters describing their distribution on the sky.

%%%%%%%%%%%%%%%%
\begin{figure}
\vspace{-2.0cm}
\centering
\includegraphics[width=8.5cm]{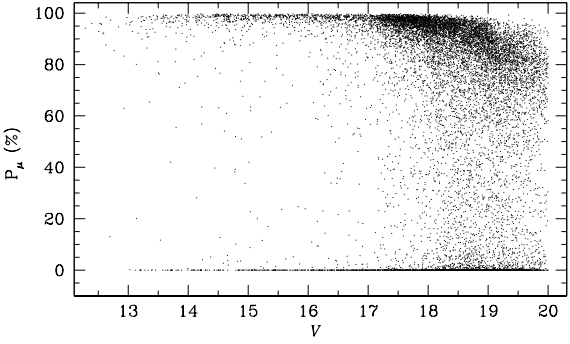}
\caption{Membership probability P$_{\mu}(\%)$ as a function  of the  $V$ magnitude
for all stars in our catalogue. At $V \sim 18.5$ mag and fainter, P$_{\mu}$
diminishes
as a result of increasing errors in the PMs.}
\label{VvsMP}
\end{figure}
%%%%%%%%%%%%%%%%

The VPD in the top left panel of Fig. \ref{cmd_inst} shows the two populations.
A tight clump at $\mu_{\alpha}$cos{$\delta$}=$\mu_{\delta}$=0.0 mas yr$^{-1}$ represents the cluster stars and a broad distribution of field stars
is centred around ($-$1.71, 4.14) mas yr$^{-1}$.
To determine the membership probability, we adopted the method
described in Balaguer-Nunez et al. (1998). This method has already
been used for $\omega$ Centauri (Bellini et al. 2009). According to this method,
first we constructed the frequency distribution of cluster stars
($\phi_c^{\nu}$) and field stars ($\phi_f^{\nu}$). The frequency
function for the $i^{\rm th}$ star of a cluster can be written as follows:\\

%\begin{equation}
    $\phi_c^{\nu} =\frac{1}{2\pi\sqrt{{(\sigma_c^2 + \epsilon_{xi}^2 )} {(\sigma_c^2 + \epsilon_{yi}^2 )}}} exp\{{-\frac{1}{2}[\frac{(\mu_{xi} - \mu_{xc})^2}{\sigma_c^2 + \epsilon_{xi}^2 } + \frac{(\mu_{yi} - \mu_{yc})^2}{\sigma_c^2 + \epsilon_{yi}^2}] }\}$, \\
%\end{equation}
where $\mu_{xi}$ and $\mu_{yi}$ are the proper motions of the
$i^{\rm th}$ star while $\mu_{xc}$ and $\mu_{yc}$ are the cluster's proper motion
centre. $\sigma_c$ is the intrinsic proper motion dispersion of
cluster member stars and ($\epsilon_{xi}, \epsilon_{yi}$) are the
observed errors in the proper motion components of $i^{th}$ star. The
frequency distribution for $i^{\rm th}$ field star is as follows:\\

%$\phi_f^{\nu} =\frac{1}{2\pi\sqrt{(1-\gamma^2)}\sqrt{{(\sigma_{xf}^2 + \epsilon_{xi}^2 )} {(\sigma_{yf}^2 + \epsilon_{yi}^2 )}}} exp\{{-\frac{1}{2(1-\gamma^2)}[\frac{(\mu_{xi} - \mu_{xf})^2}{\sigma_{xf}^2 + \epsilon_{xi}^2 } 

%$-\frac{2\gamma(\mu_{xi} - \mu_{xf})(\mu_{yi} - \mu_{yf})} {\sqrt{(\sigma_{xf}^2 + \epsilon_{xi}^2 ) (\sigma_{yf}^2 + \epsilon_{yi}^2 )}} + \frac{(\mu_{yi} - \mu_{yf})^2}{\sigma_{yf}^2 + \epsilon_{yi}^2}]\}$,  
$$
\Phi_f^{\nu}=\frac{
\exp \left\{-\frac{1}{2(1-\gamma^2)}\cdot
\left[
\frac{(\mu_{x_i}-\mu_{x_f})^2}{\sigma^2_{x_f}+\epsilon^2_{x_i}}
-\frac{2\gamma      (\mu_{x_i}-\mu_{x_f})       (\mu_{y_i}-\mu_{y_f})}
{(\sigma^2_{x_f}+\epsilon^2_{x_i})^{1/2}
(\sigma^2_{y_f}+\epsilon^2_{y_i})^{1/2}} +
\frac{(\mu_{y_i}-\mu_{y_f})^2}{\sigma^2_{y_f}+\epsilon^2_{y_i}}\right]\right\}
}     {2\pi (1-\gamma^2)^{1/2} (\sigma^2_{x_f}+\epsilon^2_{x_i})^{1/2}
(\sigma^2_{y_f}+\epsilon^2_{y_i})^{1/2}}, $$
where $\mu_{xi}$ and $\mu_{yi}$ are the proper motions of $i^{th}$ star while $\mu_{xf}$ and 
$\mu_{yf}$ are the field proper motion centre. $\epsilon_{xi}$ and $\epsilon_{yi}$ are the observed 
errors in proper motions component and $\sigma_{xf}$ and $\sigma_{yf}$ are the field 
intrinsic proper motion dispersions and $\gamma$ is the correlation coefficient. $\gamma$ 
can be calculated as\\

\begin{center}
$\gamma = \frac{(\mu_{xi} - \mu_{xf})(\mu_{yi} - \mu_{yf})}{\sigma_{xf}\sigma_{yf}}$,
\end{center}

Owing to the small observed field of the cluster we did not consider the
spatial distribution of the stars. To define the distribution function
$\phi_c^\nu$ and $\phi_f^\nu$, we used stars with a proper motion error
better than 6.5 mas yr$^{-1}$.
As expected, in the VPD, the centre of cluster stars is found to be at $\mu_{xc}$ = 0.0
mas yr$^{-1}$ and $\mu_{yc}$ = 0.0 mas yr$^{-1}$. 
Our proper motion data set could not determine intrinsic proper
 motion dispersion ($\sigma_c$) for cluster stars. Therefore, we
 adopted $\sigma_c$ = 0.2 mas yr$^{-1}$ as calculated in Sec. \ref{ppm}. For
 field stars, we have $\mu_{xf}$ = $-$1.71 mas yr$^{-1}$, $\mu_{yf}$ = 4.14 mas yr$^{-1}$,
 $\sigma_{xf}$ = 4.56 mas yr$^{-1}$  and $\sigma_{yf}$ = 4.37 mas yr$^{-1}$.  \\

The distribution of all the stars can be calculated as\\
\\
~~~~~~~~~~~~~~~~~~~~~~~~~~~~~$\phi = (n_{c}~.~\phi_c^{\nu}) + (n_f~.~\phi_f^{\nu})$,  \\

where $n_{c}$ and $n_{f}$ are the normalized number of stars for cluster and field 
($n_c + n_f = 1$). Therefore, the membership probability for $i^{th}$ star is \\
\\
~~~~~~~~~~~~~~~~~~~~~~~~~~~~$P_{\mu}(i) = \frac{\phi_{c}(i)}{\phi(i)}$, \\

A good indicator of cluster and field separation is the membership probability. 
It is plotted as a function of magnitude in Fig. \ref{VvsMP}. As seen in this plot, 
high membership probability (P$_{\mu} > 90\%$) extend down to $V \sim 19$ mag. 
At fainter magnitudes the membership probability gradually 
decreases.

\begin{figure}
\centering
\includegraphics[width=8.5cm]{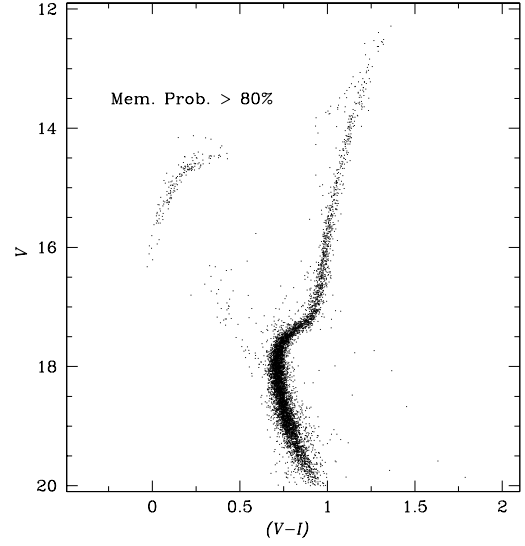}
\caption{Colour-magnitude diagram 
for the stars with membership probability $>$ 80\% }
\label{mp_cmd}
\end{figure}

Fig. \ref{mp_cmd} shows the CMD with membership probability P$_{\mu} > 80\%$. 
This CMD shows a clean main-sequence down to $V \sim 20$ mag and distinct 
population of sub-giants, red giants, horizontal branch stars, and blue 
stragglers.  

%%%%%%%%%%%%%%%%
%%%%%%%%%%%%%%%%
%____________________

%________________________________________________________________
\section{Applications}
\label{app}

\subsection{Membership probability of variables in NGC~6809}

Our catalogue can be used to assign membership probability of many variables stars 
in this cluster. Olech et al. (1999) have found 15 RR Lyrae photometric variables 
using the data taken with 1.0 m Swope telescope at Las Campanas Observatory. One
star is found in our catalogue. Its membership probability is P$_\mu=77\%$ and it is
listed in Table \ref{vari}. Our P$_\mu$ value shows that this star is a cluster member. 
NGC 6809 is 
known to have a large population of SX Phoenicis (SX Phe) variables, which are found 
in the BSS region shown in Pych et al. (2001). Out of 27, four are found in our catalogue 
and their membership probabilities are also listed in Table \ref{vari}. All four of the
SX Phe stars have a membership probability higher than 96\% and are confirmed cluster 
members. 
Bassa et al. (2008) found 31 $X$-ray sources in the direction of NGC 6809. 
Out of 31 sources, 23 are found in our catalogue and are listed in Table \ref{match_bs}.
Out of these 23 sources, 18 have a membership probability higher than 68.27\%. These 18 
$X$-ray sources may be cluster members. $X$-ray sources CX2, CX17, CX23, 
and CX26 have membership probability
less than 20\%. Hence, they may not be cluster members. Membership probability of 
CX17, CX23, and CX26 is 0\%, which means that they are field-region objects. 

\begin{figure}
\centering
\includegraphics[width=8.5cm]{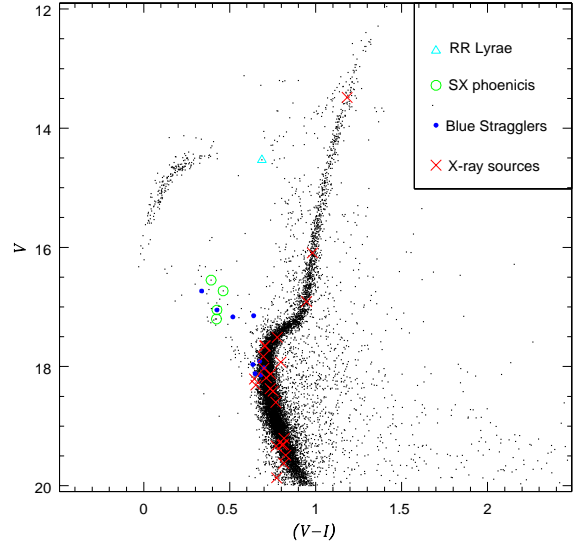}
\caption{Colour-magnitude diagram for the stars whose membership probability has been calculated.
All variables, blue stragglers, and $X$-ray sources listed in Table \ref{vari} and \ref{match_bs} 
are shown with different symbols indicated in the inset.}
\label{cmd_vari}
\end{figure}

Lanzoni et al. (2007) have studied 65 blue straggler populations by using high-resolution 
$HST$ and wide-field ground-based observations in ultraviolet and optical bands. Nine 
objects match with our catalogue and are listed in Table \ref{match_bs}. Out of these nine, 
six have P$_\mu>73\%$ and three have P$_\mu<40\%$. Hence, six blue stragglers are 
associated with the cluster and three may be field 
objects. All RR Lyrae variables, SX Phe stars, $X$-ray sources, and blue stragglers
are shown in the colour-magnitude diagram with different symbols
in  Fig. \ref{cmd_vari}.

\begin{table}  
\caption{Membership probability for RR lyrae variables provided by Olech et al. (1999) (ID$_o$) and SX 
Phoenicis variables given by Pych et al. (2001).
ID refers our catalogue number.} 
\centering
\tiny
\begin{tabular}{cccccc}
\hline
& {\small RR Lyrae}   &     &            & {\small SX Phoenicis}    & \\ 
ID$_o$  &  P$_{\mu}$   & ID        &  ID$_p$  &  P$_{\mu}$    & ID \\
&(\%)&&&(\%)&\\
\hline   

V5  &   77         &  5098     &   V24    &   96     &  9246    \\
    &              &           &   V35    &   97     & 10157    \\
    &              &           &   V38    &   97     &  5754    \\
    &              &           &   V40    &   97     &  9541    \\
\hline
\label{vari}
\end{tabular}
%\vfill
%\end{minipage}
\end{table}  

%%%%%%%%%%%%%%%%%%%%%%%%%%%%%%%%%%%%%%%%%%%%%%%%%
\begin{table}  
\caption{Membership probability for blue stragglers given by Lanzoni et al. (2001) (ID$_l$)
 and $X$-ray sources given by Bassa et al. (2008) (ID$_b$). ID refers our catalogue number.} 
\label{match_bs}
\centering
\tiny
\begin{tabular}{cccccc}
\hline
 & {\small Blue stragglers}     &           &          & {\small $X$-ray sources}          &         \\
\hfill bing \\
ID$_l$  &      P$_\mu$  &  ID  & ID$_b$  & P$_{\mu}$ &  ID     \\
&(\%)&&&(\%)&\\ 
\hline
BSS4    &   93  & 7781    &  CX2       &  13     &  8855   \\
BSS11   &   95  & 6120    &  CX3       &  91     &  7889  \\
BSS16   &   39  & 6783    &  CX4       &  96     &  7507  \\
BSS35   &   30  & 8518    &  CX5       &  86     &  8478  \\
BSS39   &   98  & 11625   &  CX6       &  98     &  3186  \\
BSS42   &   32  & 2010    &  CX7       &  94     &  4760  \\
BSS51   &   95  & 3419    &  CX8       &  93     &  4460  \\
BSS56   &   79  & 7921    &  CX9       &  95     &  7287  \\
BSS57   &   92  & 9246    &  CX11      &  74     &  6605  \\
        &       &         &  CX12      &  98     &  8594  \\
        &       &         &  CX13      &  57     &  4158  \\
        &       &         &  CX15      &  95     &  9803  \\
        &       &         &  CX16      &  94     &  7222  \\
        &       &         &  CX17      &  00     &  7125  \\
        &       &         &  CX19      &  83     &  5189  \\
        &       &         &  CX20      &  98     &  6856  \\
        &       &         &  CX21      &  83     &  2046  \\
        &       &         &  CX23      &  05     &  9983  \\
        &       &         &  CX24      &  76     &  1563  \\
        &       &         &  CX26      &  00     &  2266  \\
        &       &         &  CX27      &  96     &  1645  \\
        &       &         &  CX28      &  96     &  1117  \\
        &       &         &  CX30      &  86     &  6286  \\
\hline
\end{tabular}
\end{table}  

%%%%%%%%%%%%%%%%
%%%%%%%%%%%%%%%%

\section {The catalogue}
\label{catl}
The electronic catalogue is available at the A\&A web-site and also 
by request to the authors. In the 
catalogue, Col. (1) contains the running number; Cols. (2) and (3) provide the 
J2000 equatorial coordinates in degree, while Cols. (4) and (5) provide 
the pixel coordinates $X$ and $Y$ in a distortion-corrected reference frame. Columns 
(6) to (13) give photometric data, i.e., $U$, $B$, $V$ and $I$ magnitudes and their 
corresponding errors. If photometry in a specific band is not available, a flag 
equal to 99.999 is set for the magnitude and 0.999 for the error. Cols. (14) to 
(17) represent relative proper motions and their standard errors in mas yr$^{-1}$.
Column (18) gives
the membership probability $P_{\mu}(\%)$. The last Col. (19) gives the ID number 
of Zl11.

%________________________________________________________________
\section{Conclusions}
\label{con}

We provided a catalogue of precise proper motions and 
membership probability of stars in the wide-field region of globular cluster NGC~6809.
We have obtained precise proper motions and astrometric membership
probabilities down to  $V\sim20$ mag in 26$\times$22 arcmin$^2$ area
around the globular cluster NGC 6809. Finally, we provided the membership 
probability for different types of variable stars, blue 
stragglers, and $X$-ray sources. We also demonstrated that the CCD observations 
taken just seven years apart can provide accurate proper motions. These 
proper motions are used to separate the cluster members from field stars down to$V\sim20$ mag.

\begin{acknowledgements}

We are grateful to the referee Floor van Leeuwen for his careful reading 
of the manuscript and many useful suggestions. We also thank Dr. J. Kaluzny for 
providing us $U$ filter data for our calibration purpose. This research used
the facilities of the Canadian Astronomy Data Centre operated by the 
National Research Council of Canada with the support of the Canadian Space
Agency.
\end{acknowledgements}

%________________________________________________________________

\begin{sidewaystable*}
\centering
\caption{~The first few lines of the electronically available catalogue.} 
\vspace{0.4cm}
\centering
\tiny
%\begin{minipage}{536mm}
\begin{tabular}{ccccccccccccccccccc}
\hline\hline
ID & $\alpha_{2000}$ & $\delta_{2000}$ & $X$ & $Y$ & $U$ & $\sigma_U$ & $B$ & $\sigma_B$ & $V$ & $\sigma_V$ & $I$ & $\sigma_I$ & $\mu_{\alpha}cos\delta$ & $\sigma_{\mu_{\alpha}cos\delta}$ & $\mu_{\delta}$ & $\sigma_{\mu_{\delta}}$ &  $P_{\mu}$ & Zl11ID\\
(1)&(2)&(3)&(4)&(5)&(6)&(7)&(8)&(9)&(10)&(11)&(12)&(13)&(14)&(15)&(16)&(17)&(18)&(19)\\
$[\#]$ & $[^\circ]$ & $[^\circ]$ & [pixel] & [pixel] &[mag]& [mag] & [mag] & [mag] & [mag] & [mag] & [mag] & [mag] & [mas/yr] & [mas/yr] & [mas/yr] & [mas/yr] & $[\%]$ & $[\#]$\\
\hline
000001  &  295.013958    & -31.166472  & 2263.9486 & 236.8765 & 17.9027  &  0.0078  &      99.9990 &    0.9990 &   17.4493  &     0.0275 &   16.5824 &    0.0064  &    -0.2447  &    0.9419 &   -5.3600 &    1.1163 &  00.24  & * \\
000002  &  295.018375    & -31.165667  & 2206.5453 & 246.1274 & 99.9990  &  0.9990  &      99.9990 &    0.9990 &   19.0979  &     0.0251 &   18.1597 &    0.0381  &     5.4472  &    4.0795 &   -0.9118 &    2.8057 &  77.87  & * \\
000003  &  294.920417    & -31.169833  & 3474.9087 & 248.3862 & 99.9990  &  0.9990  &      99.9990 &    0.9990 &   18.7705  &     0.0256 &   17.8084 &    0.0482  &    -6.1344  &    0.8917 &    2.0548 &    2.2225 &  00.00  & * \\
000004  &  294.960417    & -31.167972  & 2956.8025 & 250.0339 & 99.9990  &  0.9990  &      99.9990 &    0.9990 &   17.7146  &     0.0276 &   16.3280 &    0.0117  &    -4.7667  &    2.3901 &   -7.6763 &    4.6159 &  39.61  & * \\
000005  &  294.995958    & -31.166222  & 2496.5733 & 252.6598 & 18.9460  &  0.0249  &      99.9990 &    0.9990 &   17.9888  &     0.0263 &   16.9451 &    0.0157  &    -0.2481  &    0.4291 &    2.3632 &    1.1833 &  82.53  & * \\
000006  &  295.014042    & -31.164583  & 2261.7532 & 265.3343 & 99.9990  &  0.9990  &      99.9990 &    0.9990 &   19.9458  &     0.0488 &   18.7980 &    0.0450  &    -0.4090  &    3.0839 &    4.2170 &    3.6303 &  59.11  & * \\
000007  &  294.915417    & -31.168972  & 3539.1175 & 264.6107 & 18.5690  &  0.0187  &      99.9990 &    0.9990 &   18.1759  &     0.0169 &   17.3769 &    0.0125  &     0.5766  &    0.3687 &   -6.3992 &    2.6985 &  85.33  & * \\
000008  &  294.950500    & -31.166722  & 3084.2946 & 275.5095 & 19.0454  &  0.0326  &      99.9990 &    0.9990 &   17.8176  &     0.0200 &   16.6832 &    0.0011  &    10.8374  &    2.1990 &   -0.6335 &    2.1085 &  00.12  & * \\
000009  &  294.962208    & -31.165361  & 2932.1517 & 288.2030 & 99.9990  &  0.9990  &      99.9990 &    0.9990 &   19.1391  &     0.0282 &   18.1524 &    0.0219  &     3.8851  &    2.3096 &    1.8671 &    5.2528 &  57.38  & * \\
000010  &  295.002875    & -31.163056  & 2404.7860 & 295.9780 & 99.9990  &  0.9990  &      20.6254 &    0.0489 &   19.7428  &     0.0334 &   18.6808 &    0.0234  &    -2.9163  &    3.2113 &    3.9622 &    1.4716 &  09.30  & * \\
000011  &  294.936667    & -31.165722  & 3262.0255 & 299.8488 & 17.8458  &  0.0235  &      99.9990 &    0.9990 &   17.3245  &     0.0206 &   16.4879 &    0.0097  &    -2.8828  &    1.2872 &   -0.0570 &    5.8159 &  29.44  & * \\
\hline
\label{cat}
\end{tabular}
%\end{minipage}
\end{sidewaystable*}

\end{document}